\documentclass[sigconf]{acmart}
\AtBeginDocument{%
  }

\setcopyright{none}
\acmConference{Preprint}{accepted to IVA 2025}{to appear in ACM DL.}
\acmYear{}
\acmDOI{}
\acmISBN{}

\begin{document}
\renewcommand\footnotetextcopyrightpermission[1]{}

\title{Synthetically Expressive: Evaluating gesture and voice for emotion and empathy in VR and 2D scenarios}


\author{Haoyang Du}
\authornote{Du et al. 2025. This is the author’s version of the work. It is posted here for personal use and academic dissemination. Not for redistribution. The definitive Version of Record will appear in the Proceedings of the 25th ACM International Conference on Intelligent Virtual Agents (IVA 2025).}
\affiliation{%
  \institution{Technological University Dublin}
  \city{Dublin}
  \country{Ireland}}
\email{D23128268@mytudublin.ie}

\author{Kiran Chhatre}
\affiliation{%
  \institution{KTH Royal Institute of Technology}
  \city{Stockholm}
  \country{Sweden}}
\email{chhatre@kth.se}

\author{Christopher Peters}
\affiliation{%
  \institution{KTH Royal Institute of Technology}
  \city{Stockholm}
  \country{Sweden}}
\email{chpeters@kth.se}

\author{Brian Keegan}
\affiliation{%
  \institution{Technological University Dublin}
  \city{Dublin}
  \country{Ireland}}
\email{brian.x.keegan@tudublin.ie}

\author{Rachel McDonnell}
\affiliation{%
  \institution{Trinity College Dublin}
  \city{Dublin}
  \country{Ireland}}
\email{ramcdonn@tcd.ie}

\author{Cathy Ennis}
\affiliation{%
  \institution{Maynooth University}
  \city{Kildare}
  \country{Ireland}}
\email{cathy.ennis@mu.ie}

\renewcommand{\shortauthors}{}


\begin{abstract}
The creation of virtual humans increasingly leverages automated synthesis of speech and gestures, enabling expressive, adaptable agents that effectively engage users. However, the independent development of voice and gesture generation technologies, alongside the growing popularity of virtual reality (VR), presents significant questions about the integration of these signals and their ability to convey emotional detail in immersive environments. In this paper, we evaluate the influence of real and synthetic gestures and speech, alongside varying levels of immersion (VR vs. 2D displays) and emotional contexts (positive, neutral, negative) on user perceptions. We investigate how immersion affects the perceived match between gestures and speech and the impact on key aspects of user experience, including emotional and empathetic responses and the sense of co-presence. Our findings indicate that while VR enhances the perception of natural gesture–voice pairings, it does not similarly improve synthetic ones—amplifying the perceptual gap between them. These results highlight the need to reassess gesture appropriateness and refine AI-driven synthesis for immersive environments. 

\end{abstract}


\begin{teaserfigure}
  \includegraphics[width=\textwidth]{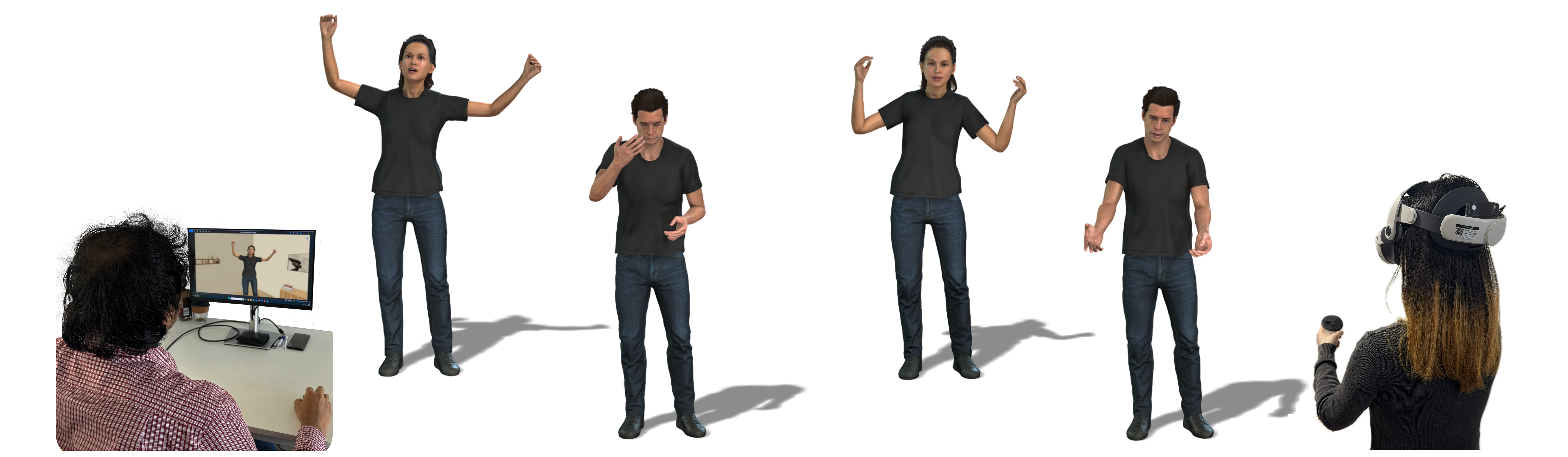}
  \caption{Participants in 2D screen (bottom left corner) VR (bottom right corner) immersion conditions. The virtual human delivers a one-minute monologue in three emotional scenarios (positive, neutral, negative) in random order, with varying gesture and voice realism. Participants then provide ratings on their experience.}
  \Description{The virtual human delivers a one-minute monologue in three emotional scenarios (positive, neutral, negative) in random order, with varying gesture and voice realism. Participants then provide ratings on their experience.}
  \label{fig:teaser}
\end{teaserfigure}



\settopmatter{printacmref=false}
\maketitle

\section{Introduction}
Advances in Artificial Intelligence (AI) and Virtual Reality (VR) technologies have profoundly reshaped computer-mediated interactions. Virtual humans — computer-generated avatars designed to engage users — are now widely utilized in customer service \cite{xu2024enabling}, education \cite{craig2018design} and healthcare \cite{rehm2016role}. Their effectiveness relies on the seamless integration of verbal and non-verbal communication channels —  gestures and speech — to build rapport, convey emotions and evoke empathetic responses \cite{qu2014conversations,volonte2020effects,danvevcek2023emotional}. Human communication is inherently multimodal, therefore, accurately modelling these behaviours is vital for creating believable virtual agents \cite{wagner2014gesture}. 

Although AI-driven co-speech gesture synthesis and modern text-to-speech (TTS) systems have become increasingly human-like, many still lack the adaptability needed to replicate diverse emotional expressions \cite{nyatsanga2023comprehensive,triantafyllopoulos2023overview}. %Consequently, users’ perceptions of authenticity may be compromised . 
In less immersive interfaces, enhancing a single modality, such as gestures or speech, improves acceptance and trustworthiness, supporting the ``maximize'' hypothesis that advocates for the highest-quality output in each channel \cite{ferstl2021human,parmar2022designing}.

However, VR’s elevated immersiveness amplifies users’ sensitivity to mismatched signals \cite{latoschik2022congruence}, owing to its potent illusions of place \cite{slater2009place} and social presence \cite{oh2018systematic}. Consistency in motion and appearance is more critical in immersive environments than in less immersive 2D interfaces, emphasizing the importance of coherent multimodal cues \cite{mal20242d}. Emotional and empathic responses, which are central to meaningful interactions \cite{paudel2024survey}, are undermined by mismatches between verbal and non-verbal empathic behaviors \cite{parmar2022designing}. Empirical evidence further highlights that congruence across channels significantly impacts emotional engagement \cite{chhatre2025evaluating,higgins2022sympathy}. These findings emphasize the importance of advancing text-to-speech and gesture integration for virtual characters while also examining how immersion affects users' sensitivity to realism. 

To address this gap in understanding how immersion impacts users' sensitivity to mismatched signals, this paper investigates the influence of different gesture-voice combinations on perceptions of virtual human interactions in both 2D and immersive VR environments. We conducted an experiment where participants watched a one-minute monologue delivered by a virtual character. The character's gestures and voice varied in realism, and were presented across different emotional scenarios in either a 2D or VR context. We find that immersive VR environments widen the perceptual gap between natural and synthetic gestures, as natural gestures become more salient while synthetic ones do not receive the same enhancement. Notably, the presence of even one natural cue—either gesture or voice—elicits emotional and empathic responses comparable to fully natural conditions, underscoring the value of partial naturalism in immersive settings.

\section{Related Work}
\textit{\textbf{Expressive Gesture Generation}}: Virtual humans often employ co-speech gestures mirroring real-human behaviours, essential for nonverbal communication to help convey ideas, and express emotions~\cite{knapp1978nonverbal}. Recent progress in datasets \cite{deichlerspatio,liu2022beat, ghorbani2023zeroeggs, kebe2024gestics,liu2024emage} has produced deep learning–based gesture generation from audio~\cite{li2021audio2gestures} text~\cite{bhattacharya2021text2gestures}, rhythm\cite{ao2022rhythmic} and contextual cues~\cite{bhattacharya2021speech2affectivegestures, voss2023augmented}. Current models are increasingly designed to learn multiple modalities or features to generate more natural and expressive gestures~\cite{zhang2024semantic, habibie2022motion, pang2023bodyformer, ao2023gesturediffuclip}, including gesture style transfer methods that aim to alter the perceived impression of gestures while preserving their semantic and temporal structure~\cite{zeng2024modifying}.  Among recent methods, emotion-conditioned gesture synthesis has received significant attention: one approach encodes emotion labels within a temporal convolutional network~\cite{liu2022beat}, while another aligns gestures with emotional beats but struggles with rare emotions~\cite{qi2024emotiongesture}. A novel line of work disentangles speech into latent vectors capturing content, style, and emotion for more flexible control~\cite{chhatre2024emotional}. These strategies may struggle with subtle expressions and require extensive data for smooth results~\cite{nyatsanga2023comprehensive}, highlighting the need to evaluate users' perception of synthesized gestures’ emotional intent in interactive contexts.

Subjective evaluation of synthetic gestures often focuses on naturalness and appropriateness through user studies in 2D environments \cite{kucherenko2023genea}, but it lacks consideration of emotional expressiveness. Gestures play a crucial role in conveying emotions, even in the absence of facial expressions \cite{ennis2013emotion}. Well-timed, spatially coherent gestures can enhance users' sense of presence, emotional engagement, and realism \cite{ennis2012perceptually}. For instance, virtual characters with high-fidelity animations have been shown to boost social presence and evoke stronger emotional responses \cite{wu2014effects}. Real-time interaction with data-driven gestures can also enhance user engagement \cite{he2022evaluating}. However, exaggerated movements combined with unsettling appearances, such as zombie-like features, can increase arousal and intensify emotions while reducing likeability, emphasizing the need to balance realism and exaggeration \cite{mousas2018effects}. Additionally, VR has become a powerful tool for emotional induction in psychological studies \cite{somarathna2022virtual}, highlighting the importance of animation accuracy for social interactions in VR \cite{ram2024virtual}. Additional research is needed to understand the emotional expressiveness of synthetic gestures in virtual environments, particularly the differences between 2D and VR contexts.

\textit{\textbf{Expressive Text-To-Speech}}: Text-to-speech (TTS) technology facilitates the conversion of written text into spoken words, with the objective of generating speech that closely approximates natural human communication \cite{taylor2009text}. Early TTS systems focused on intelligibility using articulatory, formant, and concatenative synthesis but often produced unnatural, mechanical speech, reducing user engagement \cite{cambre2020choice}. Recent advancements in Expressive Text-to-Speech (E-TTS), as detailed by Kaur and Singh \cite{kaur2023conventional}, represent a significant progression in this domain. E-TTS incorporates emotional expressiveness, speaking styles, and contextual vocal adjustments to enhance the naturalness and user experience of synthetic speech \cite{barakat2024deep}. This integration results in synthesized speech that is nearly indistinguishable from human voices. The perception of advanced neural TTS systems remains a key focus in human-computer interaction research. TTS voices have been found to be less humanlike and less trustworthy than real voices \cite{chiou2020we}. Studies comparing human voices with advanced neural TTS—using deep neural networks for natural speech—suggest that human voices are rated as more trustworthy, even though no perceptual differences in voice quality were identified \cite{do2022new}. In contrast, in VR, TTS voices are rated less appealing, expressive, and compatible with a character's appearance, reducing social presence and emotional responses \cite{higgins2022sympathy}. These findings highlight the need for further research to examine the perceptual differences of TTS voices in 2D versus VR environments.

\textit{\textbf{Mixed Realism Effects}}:
Inconsistent levels of realism in virtual human features can lead to negative impressions, commonly referred to as the uncanny valley effect \cite{macdorman2016reducing}. One study found that mismatched voice and appearance affected how anthropomorphic the agent seemed to participants \cite{alimardani2024effect}. Variations in appearance and motion have been shown to influence emotional perception \cite{mousas2018effects, amadou2023effect}. Additionally, insufficient emotional expressiveness further contributes to this phenomenon \cite{tinwell2011facial}. Researchers have explored how virtual humans expressing emotions beyond neutrality influence user perceptions, finding that high-fidelity characters displaying positive emotions are perceived as happier than low-fidelity ones, with a less pronounced effect for negative emotions \cite{higgins2021ascending}. Moreover, subtle asynchronies between verbal and non-verbal cues can disrupt emotional interpretation \cite{etienne2024systematic}. Recent findings further indicate that aligning high-fidelity facial motion with realistic appearance significantly enhances users’ empathy across a range of emotional scenarios~\cite{higgins2022sympathy,higgins2023investigating}. These findings suggest that high realism cues may enhance emotional engagement in positive scenarios, highlighting the need to examine whether gestures shape empathy when TTS and appearance realism lack impact.

Some research supports the maximization hypothesis, i.e., that each communicative channel (e.g., animation, voice) should prioritize high realism rather than maintaining uniform fidelity across all features \cite{ferstl2021human, parmar2022designing}. These studies identified TTS voice paired with robotic animation as the most mismatched combination, leading to notably low user preference \cite{ferstl2021human}. Immersion further influences virtual human perception, with VR studies revealing dynamics distinct from those observed in 2D displays. While 2D research aligns with the maximization hypothesis, VR heightens sensitivity to feature congruence, enabling congruent virtual humans to appear more plausible and less eerie \cite{mal20242d}. We anticipate that VR will enhance the impact of human gestures and speech, amplifying their perceptual distinction from AI-based gesture-voice combinations.

\textit{\textbf{Co-Presence}}: Co-presence, the feeling of sharing a virtual space with others, is fundamental to effective VR interactions \cite{schroeder2002social}. While visual realism contributes to a sense of presence, behavioural interactions play a significant role in enhancing co-presence \cite{oh2018systematic}. Studies indicate that synchronized behaviour is more impactful than visual fidelity on social presence \cite{amadou2023effect}, emphasizing the importance of well-timed and realistic behavioural cues in creating engaging experiences \cite{chu2014synchronization}. Additionally, voice realism significantly influences co-presence in VR environments. TTS voices often diminish social presence in VR, but have minimal impact in less immersive 2D settings, suggesting that VR amplifies the negative effects of TTS voices \cite{higgins2022sympathy}. These findings highlight the need to balance visual, behavioural, and auditory elements to optimize co-presence in VR environments across varying immersion levels.

\section{Study Overview}
We conducted a factorial experiment to investigate how varying forms of speech and gesture, coupled with different levels of immersion, influence participants’ perceptions of virtual characters.  Utilizing a 3 × 4 × 2 mixed factorial design (See Table S1 in the supplementary PDF and the supplementary video for study design details.), our factors included \textbf{Scenario} with three levels (Positive, Neutral, and Negative emotional contexts) as a within-subjects factor, \textbf{Gesture-Voice Combination (GVC)} with four levels (GnVn: Natural Gesture × Natural Voice; GnVs: Natural Gesture × Synthetic Voice; GsVn: Synthetic Gesture × Natural Voice; GsVs: Synthetic Gesture × Synthetic Voice) as a between-subjects factor and \textbf{Immersion} with two levels (Virtual Reality [VR] and 2D Screen-based Display [2D]) as a between-subjects factor. We present the following hypotheses:

\textbf{H1:} Increasingly synthetic voice and gesture combinations will lead to reduced perceptions of gesture-speech match, likeability, anthropomorphism, emotional responses, empathy, and co-presence.

\textbf{H2:} Higher levels of immersion will amplify the effects of H1, with VR enhancing the positive impact of natural voice and speech while further widening the differences for synthetic combinations.

\textbf{H3:}  Characters exhibiting more realistic gestures and speech will elicit higher emotional responses in positive scenarios.

\subsection{Measurements}
We employed multiple validated measures to evaluate participants' responses. Perceived match between gesture and voice was assessed using a 7-point Likert scale, following \cite{ferstl2021human}, who adapted it from the GENEA challenge's continuous scale \cite{kucherenko2020genea}. Likeability and anthropomorphism of characters were measured using items from the Godspeed questionnaire \cite{ho2010revisiting}. Co-presence was evaluated using the Networked Minds Social Presence scale \cite{harms2004internal}, focusing specifically on co-presence to capture the perceptual awareness of a virtual character's presence in our non-interactive study. Emotional responses was measured using the Affective Slider \cite{betella2016affective}, 
% a reliable tool 
for assessing valence and arousal, where participants adjusted sliders to indicate their levels for each dimension. Finally, empathy was assessed using the Measure of State Empathy \cite{powell2017situational}, which includes three self-report items for Cognitive, Affective, and Compassionate Empathy, with scores computed by summing items for each dimension as outlined in \cite{higgins2023investigating}. The specific questionnaire items are provided in Table 2 of the supplementary material.

\subsection{Participants}
We recruited 219 participants (111 female, 106 male, 2 other; aged 18–54, M = 28.9), who experienced either VR or 2D immersion with one of four gesture-voice combinations across three emotion scenarios. The VR group included 105 participants (M = 27.4, SD = 6.79), recruited via university outreach. The 2D group included 114 participants (M = 30.2, SD = 7.02), recruited through Prolific\footnote{\url{https://www.prolific.com}}, an online platform widely used in gesture research \cite{yoon2022genea, kucherenko2023genea}. To ensure consistency, recruitment was limited to participants from the same geographic region, aged between 18 and 40, and holding at least a bachelor’s degree. Additional eligibility requirements included a minimum approval rating of $ \geq 90\% $, completion of at least 100 prior studies on Prolific. We configured Prolific to prevent duplicate participation to ensure participants only viewed a single between-group condition.

\subsection{Experiment Procedures}
VR condition participants used a Meta Quest 3 headset to operate a Unity-based standalone application within a controlled laboratory setting. In the 2D condition, participants downloaded and ran the application on their computer, preserving intended quality and avoiding issues like video compression or latency. At the start of the study, participants signed a consent form and provided their demographic information. They then entered a virtual room where they received a brief training session on how to navigate the VR or 2D setup. Following the tutorial, participants observed a virtual character — either male or female — speaking to them within the virtual environment. After each scenario, participants completed a questionnaire within the same immersive setting to maintain their sense of immersion \cite{putze2020breaking}. To ensure data quality, a content-based attention check was included for 2D participants only. Those who failed were excluded from analysis. As the VR sessions were supervised in person, no attention check was required. The VR sessions lasted approximately 15–20 minutes due to the additional setup required, including fitting the headset and adjusting it for comfort. To thank them for their time, we held a raffle for two £50 vouchers for VR participants. 2D condition participants completed the task in an average of 12 minutes, as this condition did not require such preparations. For their time, 2D participants received €2 compensation, equivalent to €10 per hour. Ethical approval for the study was obtained from the institutional ethics committee.

\subsection{Stimuli Creation}
Our virtual characters, developed in Character Creator 4 (CC4)\footnote{\url{https://www.reallusion.com/character-creator/}.}, included two genders with consistent appearance to enhance variation and realism in the experiment. In Unity, we used the Standard Render Pipeline for CC4 asset compatibility and integrated CC4 shaders with basic lighting and post-processing to achieve a visually cohesive presentation. The virtual characters were positioned 1.5 meters away from participants, within the social zone \cite{sundstrom1976interpersonal}. This placement aimed to support natural conversational perception \cite{ennis2012perceptually} while ensuring gestures were clearly visible and easy to observe.

\subsubsection{Data Selection}
The animations for this study were sourced from the BEAT dataset \cite{liu2022beat}, a large-scale motion capture repository known for its detailed audio, text, and emotion annotations. We reviewed actor performances in the dataset to ensure balanced stimuli, selecting based on: (i) gesture clarity, (ii) speech clarity, (iii) appropriate emotional expression, (iv) duration of approximately one minute, and (v) unaffected dialogue context. Each scenario we selected is a one-minute monologue, providing sufficient time for participants to form perceptions of the gestures, voice, and emotional expressions. To represent positive, neutral and negative emotions, we ensured the two gender-specific actors exhibited comparable intensity. Ultimately, three scenarios — one per emotion — featuring Wayne (male) and Ayana (female) were chosen for their alignment in style and expressiveness, meeting our criteria and enhancing study validity.

\subsubsection{Speech Conditions}
Two speech conditions were employed: natural and synthetic. Natural speech samples were obtained from the BEAT dataset, while synthetic speech was generated using ElevenLabs\footnote{\url{https://elevenlabs.io/}}, a leading AI text-to-speech system chosen for its high-quality and realistic outputs \cite{brehmer2024educators}. ElevenLabs' extensive voice library enabled the selection of voices (female: Nala, male: Luke) that closely matched the original actors in accent and tone, utilizing default settings with text directly from the BEAT dataset. 

Although we used only one TTS system, ElevenLabs was chosen for its perceptual realism and broad adoption. Prior work shows that user preferences for natural speech are consistent across TTS engines and immersion levels~\cite{chiou2020we, do2022new, higgins2022sympathy}, making it a reliable proxy for comparing synthetic and natural voices.

Synthetic speech was generated using text transcripts processed through the TTS engine with embedded prompts for pauses. Temporal misalignments caused desynchronization with gestures, which were corrected in Adobe Audition by segmenting speech based on natural pauses and adjusting timing, pace and pause durations. This ensured synchronization with gestures and preserved the original performance's timing.

\subsubsection{Gesture Conditions}
We used two motion conditions: synthetic gestures generated by AMUSE~\cite{chhatre2024emotional} and natural gestures from the BEAT Dataset. AMUSE is an emotional speech-driven animation framework based on conditional latent diffusion~\cite{rombach2022high}, capable of synthesizing gestures from raw audio with explicit emotion control. We selected AMUSE due to its state-of-the-art emotion-aware performance and flexible editing features, which separate speech-driven content, emotion, and style into three latent vectors. Although only one model was employed, prior research indicates that perceptual differences largely stem from immersion context rather than the specific gesture generation method~\cite{deichler2024gesture}. Hence, AMUSE adequately serves as a representative tool for examining the impact of natural versus synthetic gestures on user perception.

In our implementation, a vision transformer (ViT)~\cite{Dosovitskiy2020Image} extracts audio features for the latent diffusion model, enabling content from one audio source (\(a_1\)) to be merged with the emotion and style from another (\(a_2\)). Specifically, we generated happy, sad and neutral gestures by combining Wayne and Ayana’s speech with emotional profiles from Scott, Solomon and Miranda respectively. The resulting gestures were imported into Blender via the SMPL-X add-on~\cite{SMPL-X:2019}, exported as BVH and retargeted in Auto-Rig Pro\footnote{\url{https://lucky3d.fr/auto-rig-pro/doc/auto_rig.html}} to a CC4-based character in iClone. We then stitched 10-second AMUSE segments to form a continuous motion sequence. Finally, to keep consistency with AMUSE-generated gestures, we adapted the natural gestures from the BEAT dataset by removing their  lower-body and finger motions to ensure consistency between conditions.

\subsubsection{Lip Synchronization and Eye Blinks}
Facial animations from the BEAT dataset were excluded because they were synchronized with recorded speech, leading to mismatches with our TTS-generated audio and reducing realism. Using a speech-to-face model could introduce artifacts and distract from our focus on gesture and speech. Consistent with prior gesture evaluation protocols \cite{kucherenko2023genea}, we opted for minimal facial animation, limited to lip synchronization and eye blinks, to avoid influencing participant judgments. Lip synchronization was achieved using AccuLIPS within iClone and enhanced with text transcripts to improve alignment accuracy. Eye-blinking was manually animated based on established conversational Spontaneous Eye-Blink Rates (SEBR) range (10.5–32.5 blinks/min) \cite{doughty2001consideration}, ensuring realistic and consistent visual behaviour across stimuli.

\subsubsection{Building VR and 2D Projects}
Our study utilized two immersion conditions, VR and 2D, with separate Unity versions to ensure consistent design, rendering and experimental setups. In both conditions, participants did not have a virtual presence and could not move within the environment: VR participants observed the character while standing in a fixed position, and 2D participants experienced a fixed camera view. Eye level for all participants was standardized and consistent across both conditions, as were rendering and performance settings. A 90 FPS frame limit was applied to both VR and 2D standalone exports, and lighting was pre-baked with no real-time lighting. Scenarios were randomized in both conditions to avoid order effects.

For the VR condition, OpenXR and Unity’s XR Interaction Toolkit were employed. Since participants could naturally rotate or tilt their heads, A `look-at' function was implemented, allowing the character to track the participant's head position.  A VR questionnaire was embedded to maintain immersion. In the 2D condition, the camera’s Field of View (FOV) and positioning were calibrated to replicate the VR perspective, ensuring consistent gesture visibility. Although screen sizes varied for prolific participants, the character remained centered with a constant FOV, ensuring consistent appearance and position while only affecting the periphery of the virtual environment. Here, the character faced the fixed camera, removing the need for a dynamic "look-at" function. An online questionnaire was embedded to collect responses (See supplementary video).

\section{Results}
The collected data was tested for normality using the Shapiro-Wilk test, which indicated deviations from a normal distribution. We therefore used the  non-parametric Aligned Rank Transform (ART) \cite{wobbrock2011aligned} for analysis of a 3×4×2 mixed ANOVA with Gesture Voice Combinations (GVC) and Immersion as between-subject factors. Post-hoc analyses used Bonferroni-corrected ART-C contrasts \cite{elkin2021aligned}.

\begin{figure} [!t]
    \centering
    \includegraphics[width=1\linewidth]{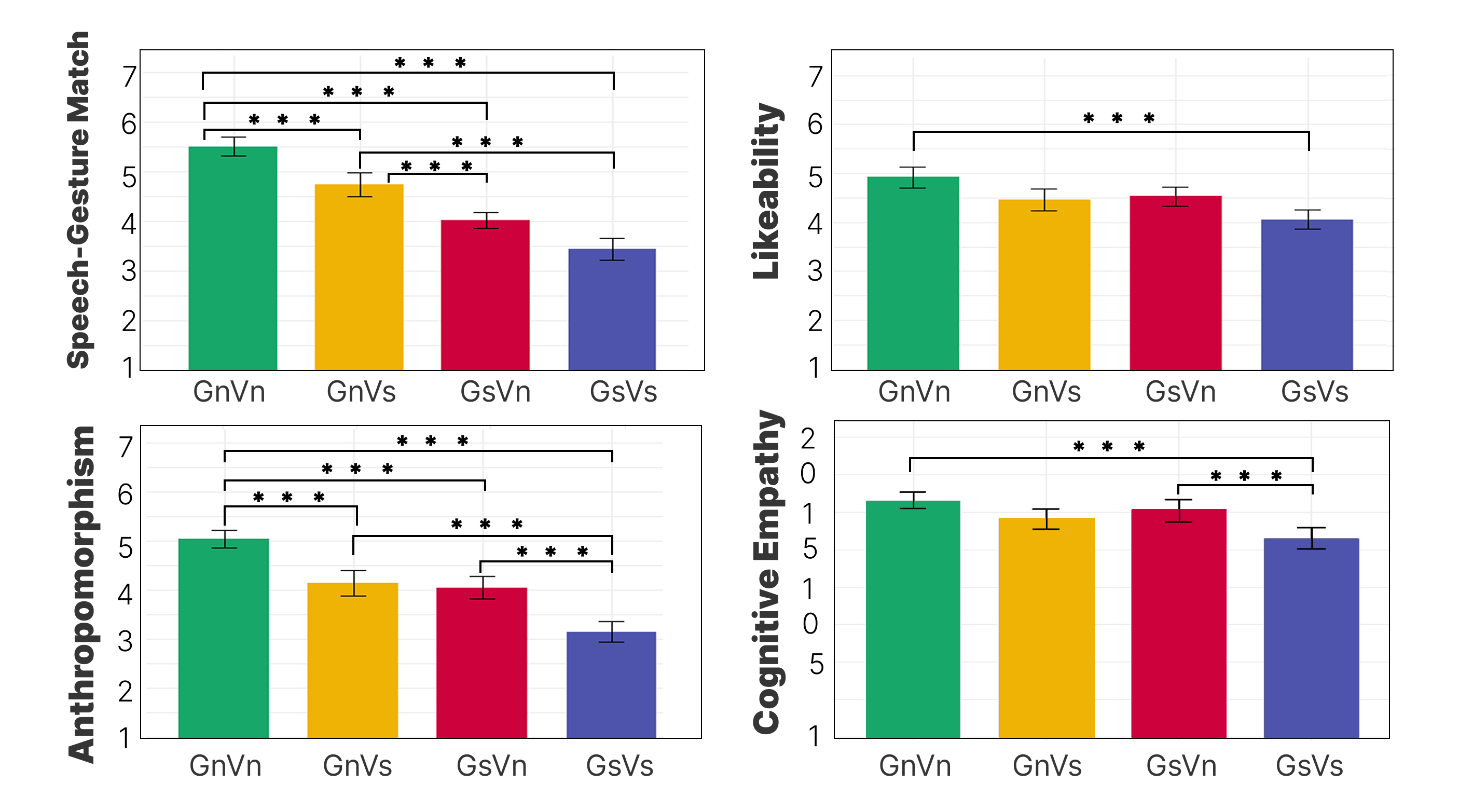}
    \caption{Main effect of GVC on various scales; fully natural combinations (GnVn) outperform synthetic combinations (GsVs) across all metrics.}
    \label{fig:MainEffect_GVC(IVA)}
\end{figure}

\begin{figure} [!t]
    \centering
    \includegraphics[width=1\linewidth]{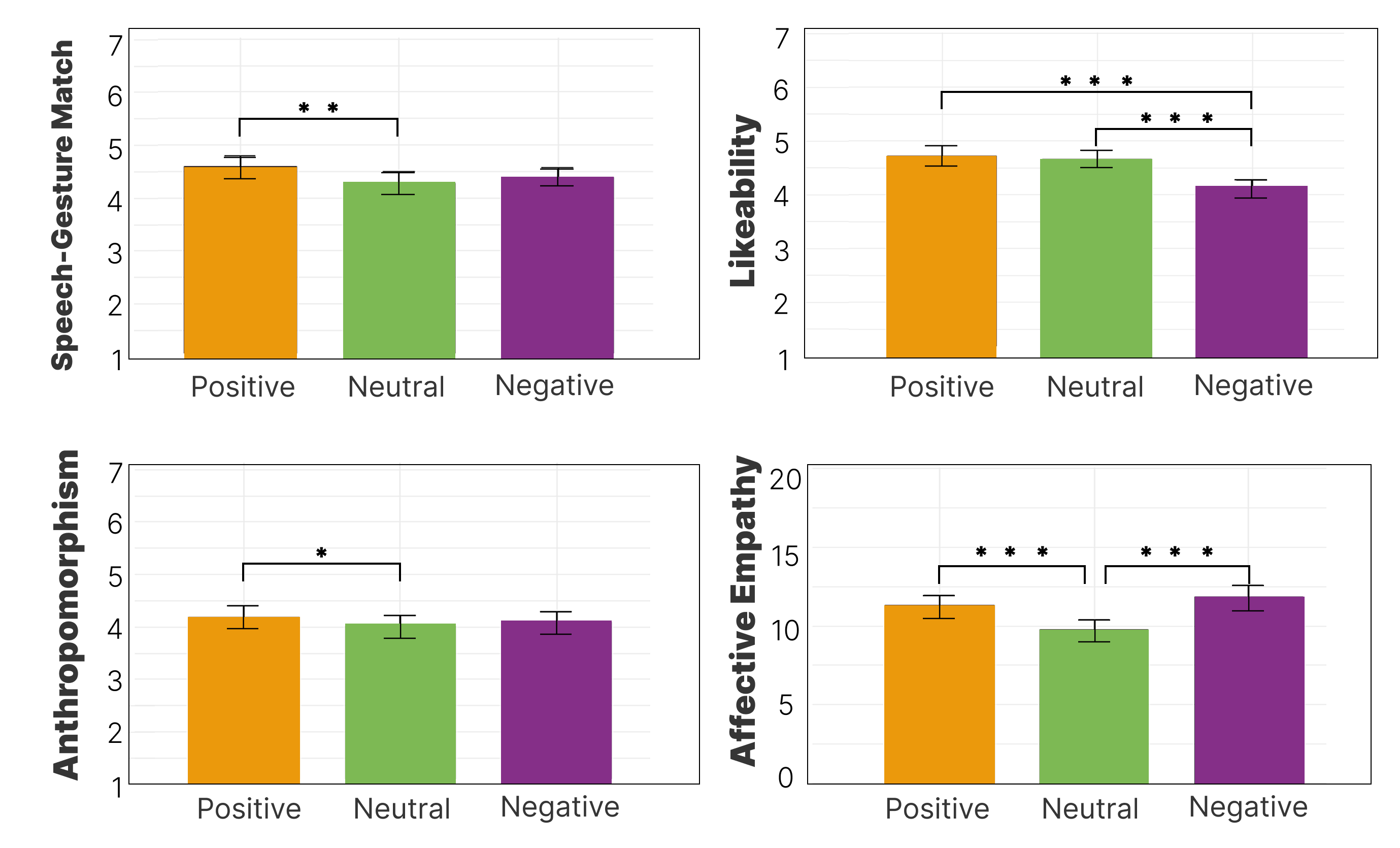}
    \caption{Main effect of Scenario on various scales: positive scenarios consistently received the highest ratings across all metrics.}
    \label{fig:Main_Effect_of_Scenario}
\end{figure}

\paragraph{\textbf{Speech-Gesture Match:}} Our analysis revealed a significant main effect of GVC (\textit{F}(3, 211) = 57.43, \textit{p} < .001) for Speech-Gesture Match , as illustrated in Fig.\ref{fig:MainEffect_GVC(IVA)}. Post hoc analysis showed that the GnVn condition produced higher match scores than all other conditions (${p} < .001.$ for all). Additionally, the GnVs condition outperformed both the GsVn (${p} < .001$) and GsVs (${p} < .001$) conditions, while no significant difference was observed between GsVn and GsVs. Immersion also had a significant effect (\textit{F}(1, 211) = 19.96, \textit{p} < .001), with VR resulting in perceived speech-gesture higher match scores than the 2D (${p} < .001$). Furthermore, Scenario had a significant effect (\textit{F}(2, 422) = 5.18, \textit{p} < .01), with positive scenarios achieving better match scores than neutral scenarios (\textit{p} < .01) (see Fig.\ref{fig:Main_Effect_of_Scenario}). 

\begin{figure} [!t]
    \centering
    \includegraphics[width=1\linewidth]{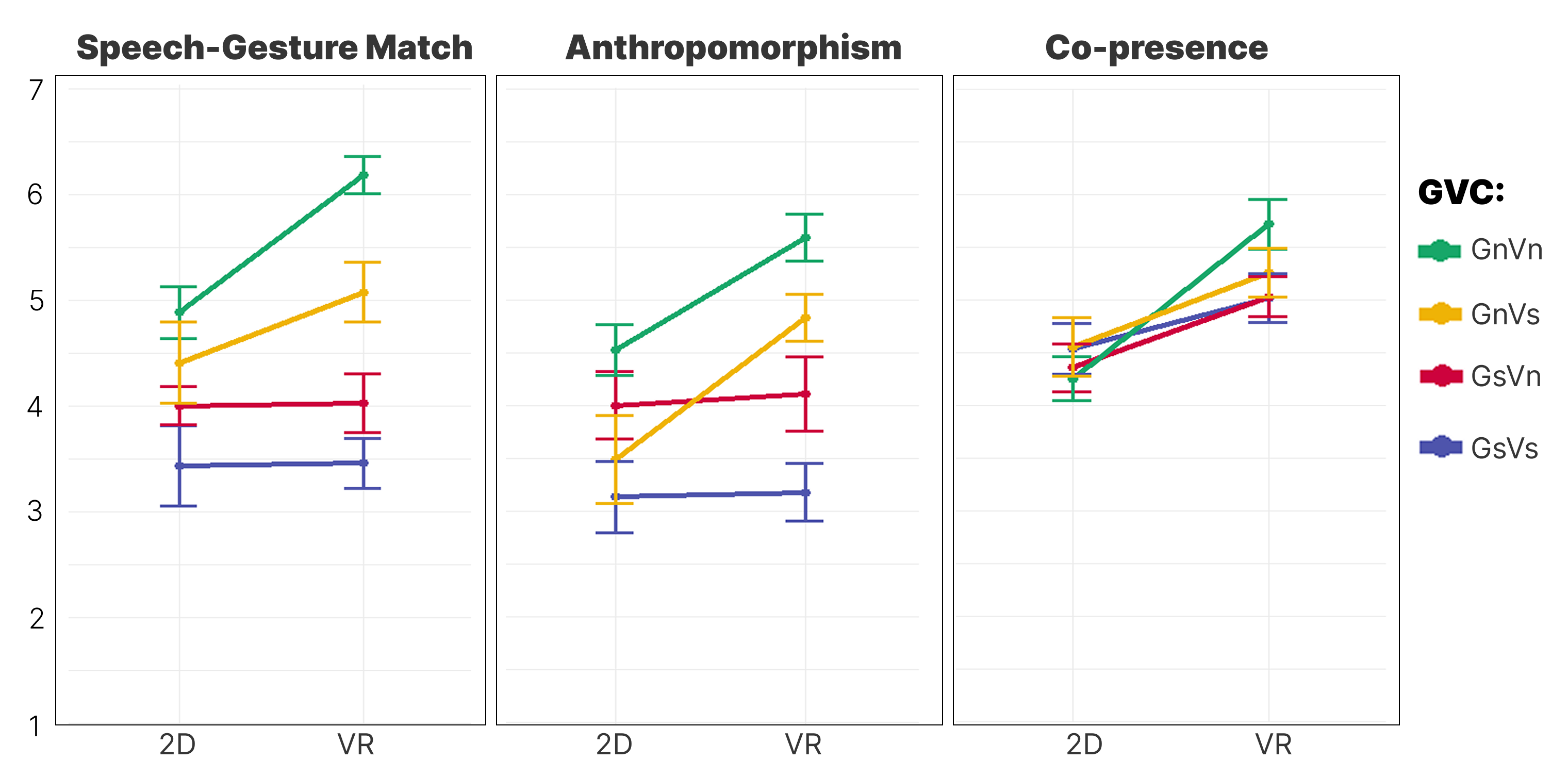}
    \caption{Interaction effect betwen GVC and Immersion on various scales; VR enhances natural combinations (GnVn) and accentuates differences with synthetic ones involving synthetic gestures.}
    \label{fig:GVC_Immersion_Interaction(IVA)}
\end{figure}

A significant interaction effect between GVC and Immersion was observed (${F}(3, 211) = 7.47, p < 0.001$), as shown in Fig.\ref{fig:GVC_Immersion_Interaction(IVA)}. The GnVn condition was perceived as significantly better matched in VR compared to 2D ($p < .001$). In both 2D and VR, GnVn differed significantly from GsVs and GsVn ($p < .001$ for all), with large effect size in VR ($r = .51$ for GsVn; $r = .61$ for GsVs). While no significant difference was found between GnVn and GnVs in 2D, GnVn scored significantly higher than GnVs in VR  ($p < .001$). For synthetic combinations, VR slightly amplified perceived match for GnVs, with a significant distinction between GnVs and GsVs in both 2D and VR ($p < .001$ for all) and a increased effect size in VR ($r = .24$ for 2D, $r = .42$ for VR). In 2D, GnVs and GsVn did not differ significantly, but in VR, GnVs outperformed GsVn ($p < .01$).

\paragraph{\textbf{Likeability:}}
The analysis revealed three significant main effects. First, GVC significantly influenced likeability (${F}(3, 211) = 6.82$, $p < .001$), with GnVn yielding higher likeability scores compared to GsVs ($p < .001$), as shown in Fig.\ref{fig:MainEffect_GVC(IVA)}. Immersion also demonstrated a significant main effect (${F}(1, 211) =12.70$, $p < .001$),  with participants in the VR condition reporting higher likeability than those in the 2D condition (${p} < .001$).  Furthermore, the Scenario factor was significant (${F}(2, 422) =24.62$, $p < .001$), with negative scenarios rated as less likeable than neutral scenario (${p} < .001$) and positive scenario (${p} < .001$), as illustrated in Fig.\ref{fig:Main_Effect_of_Scenario}. 

%This indicates that characters displaying negative emotions were the least likeable, regardless of the gesture-voice combination or immersion level.

\paragraph{\textbf{Anthropomorphism:}}
We saw three significant main effects. GVC significantly influenced Anthropomorphism (${F}(3, 211) = 22.73$, ${p} < .001$), with GnVn producing higher Anthropomorphism scores than all synthetic combinations (All: ${p} < .001$) (see Fig.\ref{fig:MainEffect_GVC(IVA)}). Additionally, GsVn and GnVs were rated significantly more human-like than GsVs (All: ${p} < .001$). Immersion demonstrated a significant main effect (${F}(1, 211) = 17.00$, ${p} < .001$), with VR resulting in higher Anthropomorphism scores than the 2D immersion (${p} < .001$). Furthermore, the scenario factor was significant  (${F}(2, 422) = 3.09$, ${p} < .05$), with positive scenarios receiving higher Anthropomorphism ratings than neutral scenarios (${p} < .05$), as shown in Fig.\ref{fig:Main_Effect_of_Scenario}.

Importantly, we also found a significant interaction between GVC and Immersion (${F}(3, 211) = 3.42$, ${p} < .05$) (see Fig.\ref{fig:GVC_Immersion_Interaction(IVA)}). Post hoc contrasts indicated that participants exposed to VR rated the GnVn with higher anthropomorphism compared to the 2D condition (${p} < .05$). Specifically, GnVn differed significantly from GsVs in both 2D (${p} < .001$) and VR (${p} < .001$), with an increased effect size observed in VR ($r = .28$ for 2D, $r = .46$ for VR). GnVn also received significantly higher human-likeness ratings than GsVn in VR (${p} < .01$), something not observed in 2D. This suggests that VR’s immersive qualities make realism-related differences more pronounced. Interestingly, the GnVs (natural gesture and synthetic speech) was also positively influenced by VR, showing higher anthropomorphism ratings compared to the 2D display (${p} < .01$).

\paragraph{\textbf{Co-Presence:}} Our analysis of Co-Presence data revealed a significant effect of immersion (${F}(1, 211) = 43.48$, ${p} < .001$), consistent with our expectation that VR enhances the sense of co-presence ($p < .05$). Scenario significantly affected co-presence (${F}(2, 422) = 4.15$, $p < .05$), with positive emotion scenarios leading to higher co-presence than neutral ($p < .05$). Additionally, The interaction between GVC and Immersion was significant, ($F(3,211)=2.98$,$p< .05$) (see Fig.\ref{fig:GVC_Immersion_Interaction(IVA)}). Post-hoc comparisons revealed that VR significantly enhanced co-presence for GnVn (${p} < .001$).

\begin{figure}[t!] 
    \centering
    \includegraphics[width=0.9\linewidth]{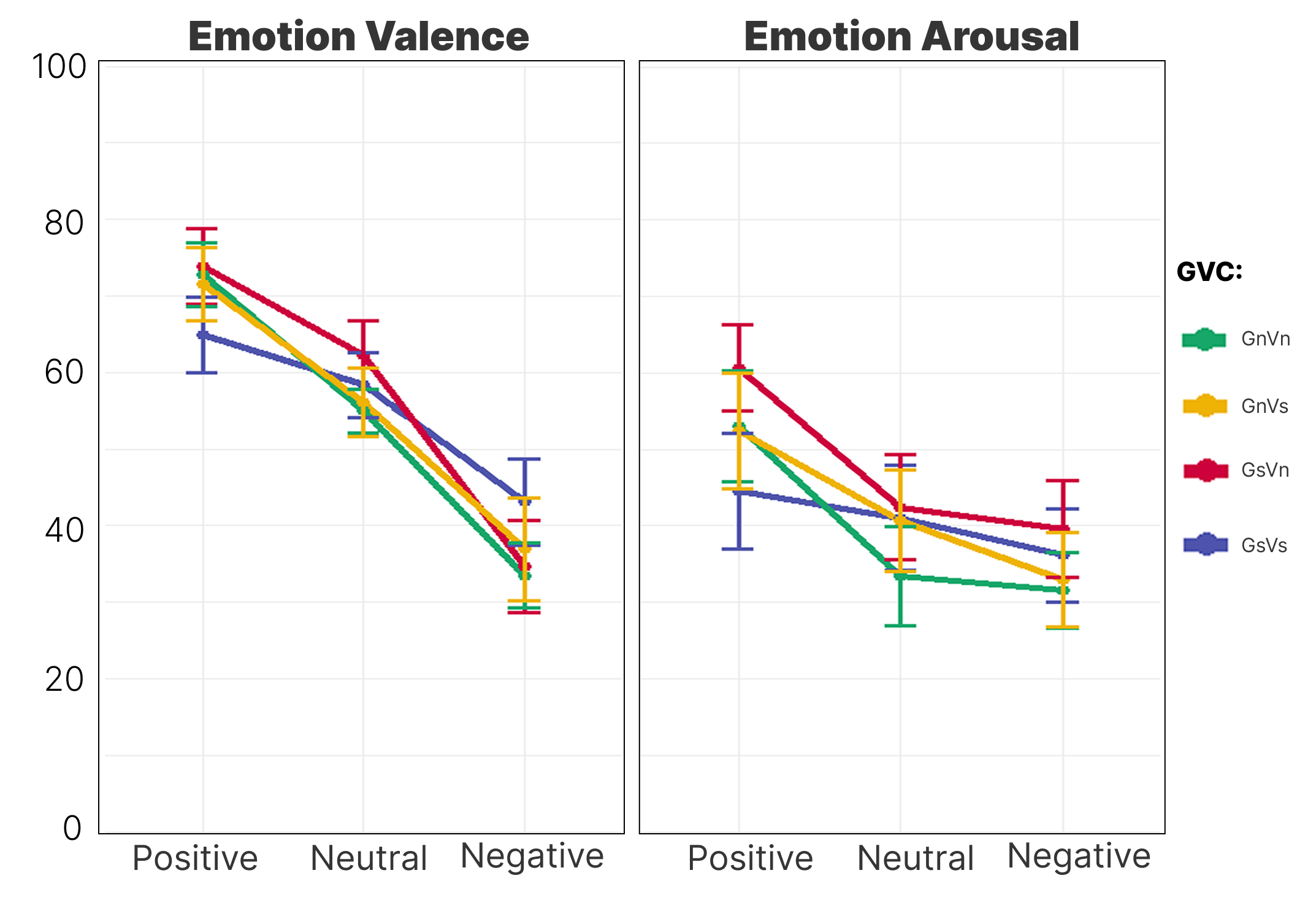}
    \caption{Interaction effect between GVC and Scenario on Emotion Engagement: the fully synthetic combination failed to differentiate positive and neutral valence and showed no arousal differences across scenarios.}
    \label{fig:GVC_Scenario_onEmotion}
\end{figure}

\paragraph{\textbf{Emotional Responses:}} We did not find a significant main effect of GVC on either valence or arousal. A significant main effect of immersion on valence was found (${F}(1, 211) = 18.10$, $p < .001$), with VR eliciting higher positive emotion than 2D ($p < .001$). Additionally, a significant interaction emerged between GVC and Scenario for both emotion valence (${F}(6, 422) = 3.33$, $p < .001$) and emotional arousal (${F}(6, 422) = 2.29$, $p < .05$), as displayed in Fig.\ref{fig:GVC_Scenario_onEmotion}. Post hoc comparisons showed that in GnVn, GsVn, and GnVs, Positive scenarios elicited higher valence than Neutral and Negative scenarios (${p} < .01$ for all). In GsVs, Positive scenarios had higher valence than Negative (${p} < .001$) but did not differ from Neutral. Additionally, arousal in GsVs remained consistent across all scenarios.

\paragraph{\textbf{Empathy Metrics:}} We found a main effect of GVC on Cognitive Empathy only (${F}(3, 211) = 5.52$, $p < .01$) (see Fig.\ref{fig:MainEffect_GVC(IVA)}). Pairwise comparisons indicated that synthetic gestures paired with natural voice (GnVn and GsVn) elicited higher empathy than the fully synthetic GsVs (GnVn:${p} < .01$, GsVn: ${p} < .05$). Additionally, immersion had a significant main effect across all empathy dimensions (Cognitive: ${F}(1, 211) = 9.65$, $p < .01$; Affective: ${F}(1, 211) = 4.12$, $p < .05$; Compassionate: ${F}(1, 211) = 14.55$, $p < .001$), indicating that VR significantly enhances participants' empathy toward characters. The scenario exhibited a significant main effect across all dimensions of empathy (${p} < .01$ for all), consistent with findings from previous research \cite{higgins2023investigating}.

\begin{figure}
    \centering
    \includegraphics[width=0.9\linewidth]{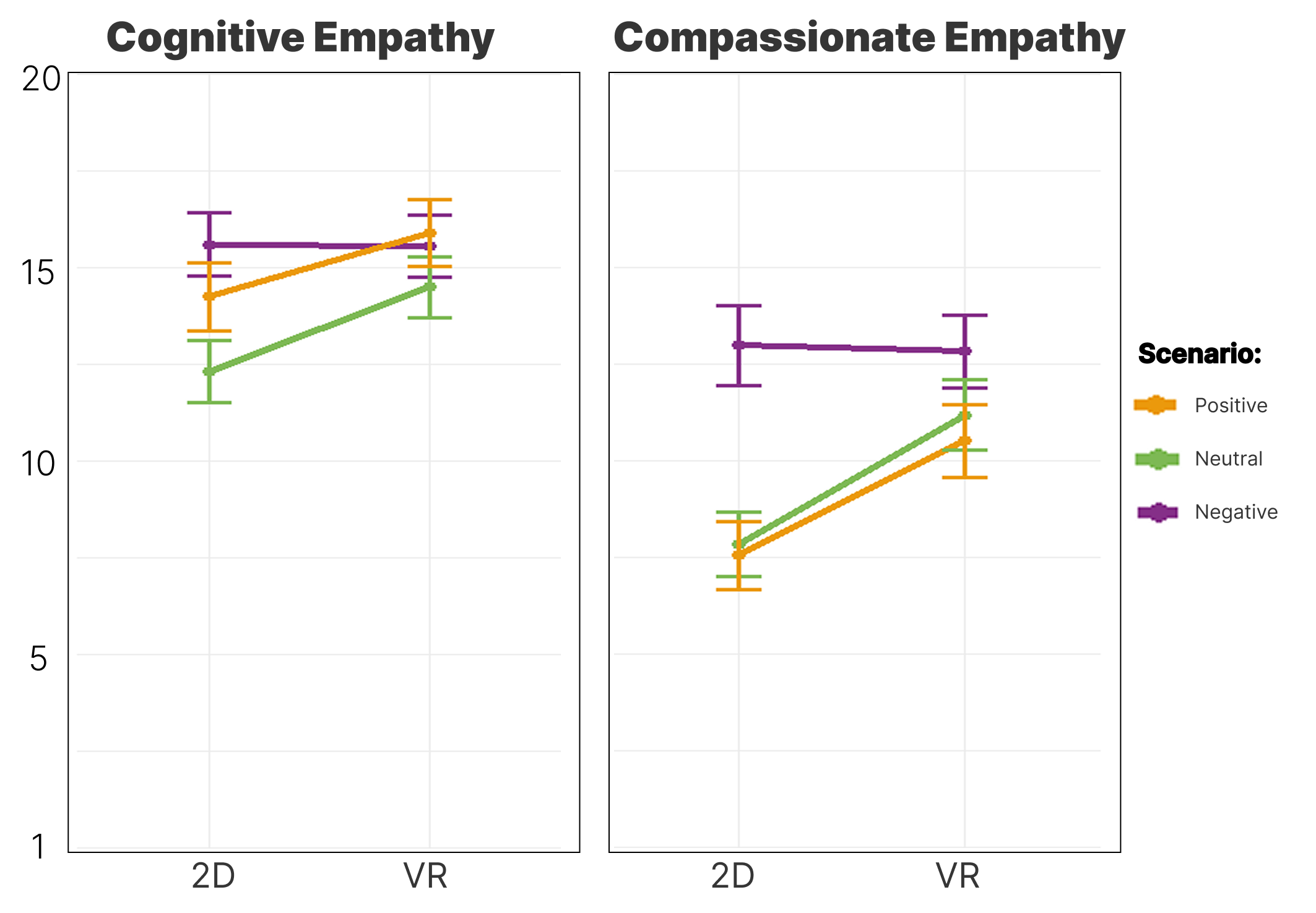}
    \caption{Interaction effect between Immersion and Scenario on Empathy: VR boosts empathy, except cognitive and compassionate empathy in negative scenarios.}
    \label{fig:Immersion_Scenario_onEmpathy}
\end{figure}

A significant interaction effect between immersion and scenario was observed for both cognitive empathy (${F}(2, 422) = 8.81$, $p < .001$) and compassionate empathy (${F}(2, 422) = 18.16$, $p < .001$), as visualized in Fig.\ref{fig:Immersion_Scenario_onEmpathy}. In VR, both cognitive and compassionate empathy scores were significantly higher than in 2D for positive and neutral scenarios (${p} < .05$ to ${p} < .001$), while no significant differences were observed in negative scenarios.

\section{Discussion}
Overall, our results support the central premise that synthetic modalities elicit dampened user responses compared to natural, motion-captured gestures and recorded human speech, and that VR intensifies these perceptions more than traditional 2D displays.

\paragraph{\textbf{Speech–Gesture Match, Likeability, and Anthropomorphism:}} 
We observed similar patterns across all measures. Essentially, using both natural gestures and natural speech (GnVn) produced the highest ratings, while fully synthetic elements consistently lowered these perceptions. These results support H1 and corroborate findings by \citet{ferstl2021human}. VR further boosted likeability across all gesture–voice combinations, contrasting with the work of Adkins et al. \shortcite{adkins2023important}, who reported minimal VR benefits in the absence of audio and with masked facial expressions, and observed no significant effects on social presence. In our study, however, the inclusion of extra audio information and contextually rich cues (e.g., emotional expressions) possibly heightens the sense of physical presence and engagement in VR.

Importantly, immersion level significantly influenced perceived congruence and anthropomorphism. VR heightened the perceived compatibility and human-likeness of fully natural (GnVn) combinations, while magnifying the differences when synthetic gestures were used, aligning with H2 and prior work \cite{mal20242d}. This suggests that greater spatial presence in VR may increase sensitivity to gesture quality, thus accentuating contrasts between natural-human and synthetic gestures. By contrast, the negative impact of synthetic voice diminished when paired with a natural-human gesture in VR. In these cases, sufficiently realistic visual cues appeared to compensate for the speech–gesture mismatch, consistent with the brain’s tendency to prioritize visual information \cite{stokes2014dominance}.

Our findings show that VR increases sensitivity to synthetic gestures, contrary to the study by \citet{deichler2024gesture}. This difference may be attributed to their use of non-human avatars, limited emotional expression, and a small sample size. Although modern deep learning-based methods can generate highly realistic gestures \cite{nyatsanga2023comprehensive}, their evaluations predominantly occur in 2D, where performance is compared to human “ground truth” \cite{kucherenko2023genea}. In extended reality (XR), congruence and plausibility remain critical \cite{latoschik2022congruence}, as mismatches in realism—especially among groups of virtual characters—can negatively affect user immersion \cite{mal2024odd}. Consequently, improving synthetic gesture quality in VR is paramount to achieving a more unified and immersive experience.

Lastly, positive scenarios garnered higher likeability scores than neutral or negative ones, reflecting a clear rank in user preference and corroborating findings from Higgins et al \shortcite{higgins2023investigating}. One plausible explanation for higher ratings of match and human likeness in positive scenarios is that the motion capture dataset—designed initially for gesture generation—included somewhat exaggerated gestures, which proved more effective at conveying positive affect.

\paragraph{\textbf{Co-presence:}} In line with H2, immersion in VR enhanced users' sense of co-presence. In particular, VR enhances natural gesture-voice combinations more than fully synthetic ones. This aligns with previous research linking virtual human realism in animation, appearance, and voice to greater co-presence \cite{wu2014effects,zibrek2017don,cherif2019anthropomorphic}. Contrary to prior findings that TTS voices in VR reduce presence due to mismatched realism and low appeal \cite{higgins2022sympathy}, our results show no significant differences in co-presence across gesture-voice combinations in 2D and VR. This divergence is attributed to our new finding that VR enhances the likeability of all gesture-voice combinations, supporting the established link between appeal and social presence \cite{oh2018systematic}. Finally, the positive scenario demonstrated a notable increase in co-presence, likely attributable to a higher degree of speech-gesture congruence and enhanced anthropomorphic qualities. In summary, our findings indicate that natural gesture-voice combinations substantially enhance users' sense of co-presence, especially in immersive VR environments.

\paragraph{\textbf{Emotional Responses:}} Firstly, the use of VR significantly enhanced the participants' experience of positive emotions, consistent with previous findings \cite{dirin2023influence}. This effect can be attributed to the highly immersive nature of VR, which surpasses other forms of media \cite{slater2009place}. We also found that gesture-voice combinations (GVCs) and scenarios jointly influenced emotional responses. Regarding emotional valence, only the fully synthetic combination failed to exhibit significant differentiation between positive and neutral scenarios. Similarly, the emotional arousal elicited by the fully synthetic combination did not vary significantly across scenario types. Contrary to H3, natural combinations did not evoke stronger emotional responses in positive scenarios compared to synthetic variants. Our new results suggest that synthetic combinations were perceived as more likeable and human-like in happy contexts, and these enhanced perceptions likely contributed to the stronger emotional responses observed. These results further demonstrated that only the fully synthetic pairing was less effective in evoking emotional responses, providing support for our H1. This aligns with previous findings that medium intensity gestures and moderate pitch variation were most effective in eliciting positive emotional responses in public speaking \cite{rodero2022effectiveness}, suggesting that naturalistic nonverbal cues play a critical role in emotional engagement.

%This study highlights the different roles of gesture and voice in shaping individuals' emotional responses in different scenarios respectively \cite{rodero2022effectiveness}. This differential impact may account for the potential attenuation of emotional responses observed in our study, wherein both modalities were synthesized.

These results imply that scenario design and higher immersion are key to fostering emotional engagement with virtual characters. While mixing natural and artificial elements can be effective, relying entirely on artificial cues may reduce the emotional impact. This insight is particularly valuable for immersive VR applications, as narrative-driven interactive experiences have been shown to effectively evoke emotional responses \cite{irshad2020increasing}. Given that virtual characters often serve as primary conduits for narratives and social interactions \cite{riedl2003character}, the integration of semi-synthetic or natural cues—rather than an exclusive reliance on fully synthetic designs—can significantly enhance their emotional impact and overall effectiveness.

\paragraph{\textbf{Empathy:}} We found that GnVn significantly enhances cognitive empathy, which involves understanding others' thoughts and emotions \cite{powell2017situational}, supporting H1. This effect may stem from the precise cues provided by natural gestures and voice, aligning with our new result that synthetic stimuli dampen emotional responses, potentially explaining the decrease in empathy. In contrast, affective and compassionate empathy, which rely more on emotional resonance, remain unaffected \cite{cuff2016empathy}. Notably, GnVn and GsVn outperform GsVs, with all effective combinations incorporating a natural voice. These results suggest that natural voice plays a critical role in maintaining cognitive empathy, which partially supports H1. The superior performance of GsVn in cognitive empathy tasks potentially arises from the natural voice's richer verbal and emotional cues compensating for synthetic gestures, while GnVs appears more neutral due to prioritization of visual cues. In contrast to Higgins’ findings \cite{higgins2022sympathy}, which suggested that TTS does not influence empathy, we attribute this discrepancy to the brevity of their scenario—only a few sentences long—which may have been insufficient to elicit meaningful empathetic responses regardless of voice type.

VR significantly enhances empathy across all dimensions, aligning with prior research suggesting that its heightened presence and engagement amplify empathic responses \cite{rueda2020virtual}. However, cognitive and compassionate empathy showed no increase in negative scenarios, likely due to a ceiling effect on uniformly high scores. Additionally, the influence of emotional scenarios on three-dimensional empathy aligns with findings that highlight the critical role of emotional states in shaping empathy \cite{higgins2023investigating}. VR is widely utilized by researchers as a medium to elicit empathy for various purposes, particularly by enabling embodiment to interact with virtual characters \cite{du2023towards}. While character production often relies on resource-intensive motion capture, our results suggest that synthetic audio cues can be an effective alternative under resource constraints, provided scenarios are carefully designed to maintain engagement.

\section{Limitations and future work}
This work has several noteworthy limitations. First, variation in participant hardware in the 2D condition—Prolific users employed diverse computer setups—contrasted with the standardized use of Meta Quest 3 headsets in the VR group, potentially leading to inconsistent visual and auditory experiences. Second, while our scenarios were balanced for gesture and speech clarity, their limited variety may have constrained engagement outcomes. Third, the digital characters used here did not reach maximum realism; more sophisticated models (e.g., MetaHuman\footnote{\url{https://metahuman.unrealengine.com}}) could prompt different viewer responses.

A limitation of this study is the use of a single gesture synthesis framework (AMUSE) and one TTS engine (ElevenLabs) to represent synthetic modalities. However, prior work suggests that user preferences and perceptual patterns are consistent across systems within the same modality. For example, Deichler et al.\cite{deichler2024gesture} found stable gesture evaluations across 2D and VR, and preferences for natural voices remain consistent across TTS engines\cite{chiou2020we, do2022new, higgins2022sympathy}. Nonetheless, this approach may be perceived as restrictive, and future work should examine a broader range of systems to enhance generalizability.

Another limitation is the mismatch in emotional expression: while our gesture model supported emotion editing, the TTS engine lacked emotional tone control, potentially causing vocal–gesture incongruence. Using emotion-capable TTS could better align modalities. Additionally, the fixed one-minute monologue limited real-time interaction, possibly obscuring subtler social responses.

Future investigations should seek to enhance diversity, expanding the participant pool beyond a western-university based population sample. Employing intermediate interfaces — such as a 3D-viewer that allows free movement but not full immersion — may also clarify whether graded levels of immersion produce meaningful differences in perception and co-presence. More realistic and varied digital humans, potentially with fully rigged facial expressions or alternative appearance styles, could provide deeper insights into uncanny valley thresholds and anthropomorphism. Introducing real-time user–avatar interactions, incorporating eye-tracking to measure user engagement, and exploring proxemics could more closely mimic everyday social encounters and thus yield stronger external validity. In addition, it is possible that gesture had a greater impact than natural speech because users are familiar with synthetic voices. It would be interesting to explore how familiarity with synthetic cues influences perception. We also plan to investigate the impact of individual elements of conversational gesture (e.g., gesture type, timing, activation level) to further guide the development of co-speech gesture synthesis.

\section{Conclusion} 
In summary, our findings highlight the substantial advantages of combining natural gestures and speech to enhance users’ perceptions of realism, likeability, anthropomorphism, and co-presence — benefits that are magnified in virtual reality (VR). Examining a wide range of different measures simultaneously, Our study is first to identify that the gap in perceived quality between natural and synthetic gestures becomes particularly pronounced in VR, emphasizing the need to develop more lifelike synthetic gestures for immersive applications. Moreover, our results emphasize the key role of both scenario context and immersion in driving emotional engagement and empathy. While fully synthetic cues tend to diminish emotional resonance and reduce cognitive empathy, semi-synthetic combinations can still evoke strong affective responses — especially when enhanced by natural voices that convey richer verbal and emotional information.

We present these implications to highlight how tailoring vocal and gestural cues, alongside immersive platforms, can optimize both emotional engagement and empathy in virtual scenarios. Most gesture evaluation studies to date have used limited avatar representations, scenario contexts, and immersion environments to explore the relationship between gesture and voice. Our study suggests that gesture generation researchers should note the magnified gap between synthetic and natural gestures in VR and account for it in future designs and evaluations. Additionally, a single natural cue—either in voice or gesture—is sufficient to significantly boost immersion and empathy, offering a practical direction for developers working with resource-constrained systems. As VR and AI technologies rapidly evolve, further research into how best to coordinate voice and gesture is essential for creating immersive, engaging, and emotionally resonant virtual interactions.

%\iffalse

%\section{Acknowledgments}
%This work was conducted with the financial support of the Research Ireland Centre for Research Training in Digitally-Enhanced Reality (d-real) under Grant No. 18/CRT/6224. For the purpose of Open Access, the author has applied a CC BY public copyright licence to any Author Accepted Manuscript version arising from this submission

%\fi
%\section{Appendices}

%If your work needs an appendix, add it before the
%``\verb|\end{document}|'' command at the conclusion of your source
%document.

%Start the appendix with the ``\verb|appendix|'' command:
%\begin{verbatim}
%  \appendix
%\end{verbatim}
%and note that in the appendix, sections are lettered, not
%numbered. This document has two appendices, demonstrating the section
%and subsection identification method.


\begin{acks}
This work was conducted with the financial support of the Research Ireland Centre for Research Training in Digitally-Enhanced Reality (d-real) under Grant No. 18/CRT/6224.
\end{acks}

\bibliographystyle{ACM-Reference-Format}
%%\bibliography{ref_opt}

%\appendix

%\section{Appendices}

%
%\begin{figure}[H]
 %   \centering
%    \includegraphics[width=1\linewidth]{Plot/Slider.png}
%    \caption{Emotional slider scale}
%    \label{fig:AffectiveSlider}
%\end{figure}

\end{document}